\documentclass[reprint,amsmath,amssymb,aip,eqsecnum,jcp]{revtex4-1}

\usepackage{dcolumn}
\usepackage{amssymb}
\usepackage{amsmath}
\usepackage{amsfonts}
\usepackage{amsbsy}
\usepackage{graphicx}
\usepackage{color}
\usepackage{bm}
\usepackage{bbm}

\usepackage[normalem]{ulem}

\newcommand\beq{\begin{equation}}
\newcommand\eeq{\end{equation}}
\newcommand\beqa{\begin{eqnarray}}
\newcommand\eeqa{\end{eqnarray}}
\newcommand{\non}{\nonumber\\}
\def\bal#1\eal{\begin{align}#1\end{align}}

\newcommand{\bq}{\begin{eqnarray}}
\newcommand{\eq}{\end{eqnarray}}
\newcommand{\bqn}{\begin{eqnarray*}}
\newcommand{\eqn}{\end{eqnarray*}}
\newcommand{\nn}{\mathbf{n}}
\newcommand{\rr}{\mathbf{r}}

\newcommand{\sss}{\mathbf{s}}
\newcommand{\qq}{\mathbf{q}}
\newcommand{\CL}{{\cal L}}
\newcommand{\xx}{\mathbf{x}}
\newcommand{\FF}{\mathbf{F}}
\newcommand{\zero}{{(0)}}
\newcommand{\one}{{(1)}}

\newcommand{\sz}{{S$_0$}}
\newcommand{\spl}{{S$_+$}}
\newcommand{\sm}{{S$_-$}}
\newcommand{\wsz}{{wS$_0$}}
\newcommand{\wspl}{{wS$_+$}}
\newcommand{\wsm}{{wS$_-$}}
\newcommand{\az}{{A$_0$}}
\newcommand{\aplus}{{A$_+$}}
\newcommand{\am}{{A$_-$}}
\newcommand{\waz}{{wA$_0$}}
\newcommand{\wapl}{{wA$_+$}}
\newcommand{\wam}{{wA$_-$}}

\begin{document}
\title{Depletion force in the infinite-dilution limit in a solvent of
  nonadditive hard spheres}

\author{Riccardo Fantoni}
\email{rfantoni@ts.infn.it}
\affiliation{Dipartimento di Scienze dei Materiali e Nanosistemi,
  Universit\`a Ca' Foscari Venezia, Calle Larga S. Marta DD2137,
  I-30123 Venezia, Italy}

  \author{Andr\'es Santos}
\email{andres@unex.es}
\homepage{http://www.unex.es/eweb/fisteor/andres/}

\affiliation{Departamento de F\'{\i}sica and Instituto de Computaci\'on Cient\'ifica Avanzada de Extremadura (ICCAEx), Universidad de
Extremadura, Badajoz, E-06071, Spain}

\date{\today}

\begin{abstract}
The mutual entropic depletion force felt by two solute ``big'' hard spheres immersed in a binary mixture
solvent of nonadditive ``small'' hard spheres  is calculated as a
function of the surface-to-surface distance by means of canonical Monte Carlo simulations and through a recently proposed rational-function approximation [Phys. Rev. E \textbf{84}, 041201 (2011)]. Four representative scenarios
are investigated:  symmetric solute particles and the limit
where one of the two solute spheres becomes a planar hard wall, in both cases with symmetric and
asymmetric solvents. In all cases, the influence on the
depletion force
due to the nonadditivity in the solvent is determined in the mixed
state. Comparison between results from the theoretical approximation
and from the simulation shows a good agreement for  surface-to-surface
distances greater than the smallest solvent diameter.
\end{abstract}

\maketitle
\section{Introduction}
\label{sec:introduction}

In chemical physics one often finds {solute particles immersed in {a solvent}. Even though the solute particles  interact
through a \emph{true} potential, an
important problem consists of reducing the solute-solvent}  system of particles to an
equivalent one made of only the solute particles but interacting
through an \emph{effective} potential. This problem has been much
studied for the paradigmatic case of an athermal  mixture of
additive hard spheres (AHS)\cite{DRE99a}  and for the more
general case of nonadditive hard spheres (NAHS).\cite{REA01,LR01,ELPGMC14}
The
problem is usually solved in a two-step procedure. Starting from the
pioneering work of Asakura and Oosawa,\cite{AO54} one first
determines the effective \emph{pair} potential, the so-called \emph{depletion
entropic potential}, between \emph{two} ``big'' solute hard spheres
(in three\cite{RED00,YSH08,AWRE11} or  two\cite{CRM03}
dimensions) immersed in a solvent of ``small''
hard spheres. Once this stage has been carried out, one can study the properties of a fluid of particles  interacting
with such an effective pair potential.\cite{BBF96} While the
assumption of pairwise additivity is essentially uncontrolled, since
the presence of a third particle in the vicinity of a pair of solutes
will alter the solvent (the depletant) spatial distribution, it is
expected that such limitations of the pairwise additivity approximation become progressively less relevant
on decreasing the solute density and/or the size ratio between the
diameter of a solvent particle and that of a solute particle.
The oscillations in the depletion potential, for example, are found to
be responsible for gelation in binary mixture of hard spheres\cite{GA09} and for spatial heterogeneity in bimodal colloidal suspensions.\cite{SSN11}

In the present work we are interested in the first step of such a
programme. The depletion potential problem has been studied in several
different scenarios. One can have nonspherical solute\cite{KRD06} or  solvent\cite{LTM08} particles. For
spherical solute  and  solvent particles, the case
we are interested in, the solvent itself may be an AHS mixture (binary,\cite{HTW04} multicomponent,\cite{RK06} or polydisperse\cite{M95,CNW96}). Additionally,  the
solvent particles may have various kinds of
interaction.\cite{E04,CMMNTV07,LTJ09,J09}
When the solvent particles interact with a potential which has some
attraction, an interesting issue is the one of understanding how the
depletion  or force will be affected upon
approaching the gas-liquid coexistence critical point of the solvent,
where the critical fluctuations are expected to give rise to the so-called thermodynamic Casimir forces.\cite{BBD00,HHGDB07,GZTS12,GZS12}

Recently, we constructed an approximate theory for the
structure and the thermodynamics of a general NAHS multicomponent mixture,\cite{FS11,FS13} which we
called the (first order) rational-function approximation (RFA). The theory provides a fully analytical representation of the radial distribution functions in Laplace space which extends to the nonadditive case  the exact solution of the Percus--Yevick (PY) integral equation for AHS mixtures.\cite{L64,YSH98}
It is the purpose of the present work to use the RFA theory
to predict the depletion force when the solvent is a NAHS binary mixture  and to compare our theoretical predictions
with Monte Carlo (MC) simulation  results. We clearly want to
avoid demixing\cite{BH91,SHY10} in the solvent. This
restricts the combinations of solvent density and (positive) nonadditivity that
we are allowed to choose. An interesting open problem, that we leave to
a future study, is the  study of how the depletion force is
affected by approaching the demixing critical point on the critical
isochore.

In order to find the depletion force in the simulations we followed
the MC method of Dickman et al.\cite{DAS97} In
molecular dynamics simulations, however, a different strategy\cite{AL01} is more suitable. We decided not to determine
the depletion potential from the force because the spatial integration of the latter can
introduce additional uncontrolled uncertainties. On the other hand, it is possible
to determine the depletion potential directly in a MC simulation by
allowing the two solute impurities to move.\cite{GZS12}

We will consider four different scenarios: (i) two symmetric solute particles in a
symmetric solvent, (ii) two symmetric solute particles in an asymmetric solvent, (iii)
extremely asymmetric solutes, in the limit where one of the two
solute spheres reduces to a planar hard wall,\cite{FS13}
in a symmetric solvent, and (iv) the same situation but in an asymmetric solvent.

The paper is organized as follows. In Sec.\ \ref{sec:model}, we
introduce the fluid model we set up to study further on, while in Sec.\
\ref{sec:depletion}  the observable  to be measured in
 MC simulations and estimated with our RFA theory is described. Details about our MC simulations are given in Sec.\
\ref{sec:mc}. Section
\ref{sec:results} presents the numerical and theoretical results for
the depletion force and compares them. The paper is closed in Sec.\ \ref{sec:conclusions} with some final remarks.

\section{The model}
\label{sec:model}

We consider the following general model. Two solute big
hard spheres (the impurities) of species $a$ and $b$ and diameters
$\sigma_a=\sigma_{aa}$ and $\sigma_b=\sigma_{bb}$ with
$\sigma_{ab}=\frac{1}{2}(\sigma_a+\sigma_b)$ are immersed in a NAHS binary mixture solvent made of $N_\mu$ small hard spheres
of species $\mu=1,2$ of diameter $\sigma_\mu=\sigma_{\mu\mu}$ in a
volume $V$, such that
\beq
\sigma_{12}=\frac{\sigma_1+\sigma_2}{2}(1+\Delta)
\eeq
 with $\Delta>-1$ measuring the solvent nonadditivity.
The solute-solvent interaction is assumed to be additive, i.e., $\sigma_{\mu \alpha}=\frac{1}{2}(\sigma_\mu+\sigma_\alpha)$ with $\mu=1,2$ and $\alpha=a,b$.

Without loss of generality, we take $\sigma_1(\leq\sigma_2)$ as length unit. Thus, we define the solvent/solvent size ratio
$\sigma_2/\sigma_1\geq 1$, the solute/solute size ratio $\sigma_b/\sigma_a\geq 1$, and the solute/solvent size ratio
$\sigma_a/\sigma_1>1$. The solvent total number density is
$\rho=N/V=\sum_{\mu=1}^2 N_\mu/V=\sum_{\mu=1}^2\rho_\mu$ and the mole fraction of species $\mu=1,2$ is $x_\mu=\rho_\mu/\rho$, with $x_1+x_2=1$. {}From this we can introduce  the partial packing fractions
$\eta_\mu=\frac{\pi}{6}\rho x_\mu\sigma_\mu^3$ and the nominal total
packing fraction $\eta=\sum_\mu\eta_\mu$.

The model is characterized by the following set of six independent dimensionless parameters:
$\eta$, $x_1$, $\sigma_2/\sigma_1$ and $\Delta$, defining the solvent, and $\sigma_b/\sigma_a$ and $\sigma_a/\sigma_1$,
defining the solute. Note that the model can also be  obtained from the more general
one of a quaternary  mixture with $a=3$, $b=4$ in the limit
of infinite solute dilution $x_3\to 0$, $x_4\to 0$.\cite{YSH08}

{The depletion force is formally independent of the solvent-solvent interaction (see Sec.\ \ref{sec:depletion}).\cite{A89,DAS97}  But of course it depends on the
\emph{local}  solvent  density in the neighborhood of the solute particles and such a density is affected by the solvent-solvent and solvent-solute interactions.}
A natural question then arises: As the solvent-solvent
nonadditivity is switched on, how the induced change in the local solvent density affects
the  depletion force?
Clearly, far away from the solute spheres there will be no change in
the almost constant local density, i.e., the bulk density. But the local
density in the vicinity of the solute particles would change and thereby so would
the force. To first order in density, however, the depletion force is completely independent of the solvent-solvent interaction,\cite{YSH08} so the influence of  nonadditivity is absent. Thus, one can expect the effect  to be small for dilute solvents but its impact as the bulk solvent density increases is uncertain.

We could alternatively switch on a solute-solvent nonadditivity,\cite{REA01,RE01,ELPGMC14} but
this case is somewhat less interesting than the previous one. For
example, in the case of two solute spheres of diameter $\sigma_a$ immersed in a one-component solvent of spheres
of diameter $\sigma_1$ with $\sigma_{1a}\neq \frac{1}{2}(\sigma_1+\sigma_a)$, one can  map the problem onto an additive one where the solute particles have an effective diameter $\sigma_a^\text{eff}=2\sigma_{1a}-\sigma_1$, provided that $\sigma_{1a}\geq \frac{1}{2}\sigma_1$. The
effective problem determines the depletion force for $r>\sigma_a^\text{eff}$, so that the original problem becomes completely solved in the case of negative
nonadditivity (since then $\sigma_a>\sigma_a^\text{eff}$), while in the case of positive nonadditivity  it only
remains unsolved in the region $\sigma_a<r<\sigma_a^\text{eff}$. For this reason, we
will not consider solute-solvent nonadditivity in our analysis.

In this study, we will first restrict ourselves to the particular case of
equal solute impurities ($\sigma_b/\sigma_a=1$) and consider both a symmetric
($\sigma_2/\sigma_1=1$, $x_1=\frac{1}{2}$) and an asymmetric ($\sigma_2/\sigma_1\neq 1$, $x_1\neq\frac{1}{2}$)
nonadditive solvent. Our aim is to assess in both cases  the
effect of the solvent nonadditivity on the depletion force.
Then, we will consider the case of extremely
asymmetric solute impurities in the limit  $\sigma_b/\sigma_a\to\infty$, where
one of the two impurities is seen as a hard planar wall both by the
other solute sphere and by the solvent species.

\section{The depletion force}
\label{sec:depletion}

We want to determine the force exerted on one big solute sphere
immersed in a solvent of small spheres due to the presence of a
second big solute sphere, assuming a hard-core
repulsion between the solvent and the solute. The solvent in the presence
of only one solute sphere at the origin will keep being an isotropic
fluid (even if not homogeneous anymore) and the solute sphere will
feel a zero net force. However, if we add a second solute sphere in the solvent, the isotropy symmetry will be broken (we are then left with a
solvent fluid with axial symmetry around the axis connecting the
centers of the two solute spheres) and, as a consequence, each
solute sphere will exert an effective force $\FF$ on the other one, mediated
by the solvent. This force has the form\cite{A89}
\beq
{\beta\FF(r) = -\int_S dA\, \rho^{(r)}({\mathbf{r}_s})\widehat{\nn},}
\label{FF}
\eeq
where $\beta=1/k_BT$ is the inverse temperature parameter, the integral is carried out over the surface $S$ of the sphere
centered on the solute particle experiencing the force, $dA$ is an elementary area on $S$,  $\widehat{\nn}$ is the outward normal unit vector, and $\rho^{(r)}({\mathbf{r}_s})$ is the local density of the solvent (in the presence
of the two solute spheres) {at the point ${\mathbf{r}_s}$} on the surface $S$.

\subsection{{Monte Carlo implementation}}
\subsubsection{One-component solvent}
Let us first assume a one-component solvent made of $N$ spheres of diameter $\sigma_1$ and coordinates
$\rr_i$ ($i=1,\ldots,N$) in a volume $V$. The solute particle of
species $a$ is centered at {$\rr_a$} and the solute particle of
species $b$ is centered at {$\rr_b=\rr_a+r\widehat{\rr}$.} According to Eq.\ \eqref{FF}, the force
$\FF_{ab}(r)=F_{ab}(r)\widehat{\rr}$ felt by sphere $b$ due to the presence
of  sphere $a$ is then\cite{A89}
\beq
\beta F_{ab}(r)=-\sigma_{1b}^2\int d\Omega_s\,\cos\theta_s\rho^{(r)}({\rr_b}+\sigma_{1b}\widehat{\mathbf{s}}),
\label{Fab}
\eeq
where $d\Omega_s=\sin\theta_sd\theta_sd\varphi_s$ is
the elementary solid angle spanned by $\widehat{\sss}$ taking the polar
axis along $\widehat{\rr}$,
$\rho^{(r)}(\qq)=\langle\sum_i\delta(\qq-\rr_i)\rangle$ is the local
density of the solvent in the presence of the two solute spheres at a
center-to-center distance $r$, and
$\langle\cdots\rangle$ is a thermal average.

The expression \eqref{Fab} for the
depletion force is {formally} independent of the interaction between the
solvent particles and holds as long as we have a hard-sphere
interaction between the solvent and the two solute spheres. Clearly,
due to the axial symmetry of the solvent fluid,
$\rho^{(r)}({\rr_b}+\sigma_{1b}\widehat{\mathbf{s}})=\langle\sum_i\delta(\sigma_{1b}\widehat{\mathbf{s}}-\sss_i)\rangle$, with
$\sss_i=\rr_i-{\rr_b}$, is a function of $\sigma_{1b}$ and $\theta_s$ only. Notice
that, by Newton's third law, we must have $F_{ab}=-F_{ba}$. In terms of
the potential of mean force $\beta u_{ab}(r)=-\ln g_{ab}(r)$, where
$g_{ab}(r)$ is the solute-solute radial distribution function {in the presence of the solvent}, we have
\beq \label{rfa-force}
\beta F_{ab}(r)=-\beta\frac{du_{ab}(r)}{dr}=
\frac{g^\prime_{ab}(r)}{g_{ab}(r)}.
\eeq

In MC simulations we can calculate the force by means of
\bal
\label{mc-force}
\beta F_{ab}(r)=&-\sigma_{1b}^2\left\langle\sum_i\int d\Omega_s\,\cos\theta_s
\delta(\sigma_{1b}\widehat{\sss}-\sss_i)\right\rangle\non
\approx&-{3}\sigma_{1b}^2\left\langle\sum_i
\frac{{\Pi_{s_i-\frac{\epsilon}{2},s_i+\frac{\epsilon}{2}}(\sigma_{1b})}\cos\theta_{s_i}}
{{(s_i+{\frac{\epsilon}{2}})^3-(s_i-{\frac{\epsilon}{2}})^3}}
\right\rangle,
\eal
where the boxcar function
{$\Pi_{a,b}(x)=1$ if $a\leq x<b$} and zero
otherwise, ${\epsilon}$ is a discretization of the $s$ variable, and in
the second line of Eq.\ (\ref{mc-force}) we have discretized the radial
part of the Dirac delta function. We can also rewrite
Eq.\ (\ref{mc-force}), by neglecting the term in ${\epsilon}^3$ in the
denominator, as follows
\beq
F^*_{ab}(r)\equiv\sigma_1\beta F_{ab}(r)
\approx-\sigma_1 I^{(r)}(\sigma_{1b}),
\eeq
where
$F^*_{ab}(r)$ is the dimensionless force and
\beq
I^{(r)}(s)=\left\langle\sum_i\frac{{\Pi_{s-\frac{\epsilon}{2},s+\frac{\epsilon}{2}}(s_i)}\cos\theta_{s_i}}
{{\epsilon}}\right\rangle.
\eeq
In the simulations, $I^{(r)}(s)$ is evaluated at $s=s_\kappa=\sigma_{1b}+(2\kappa+1){\epsilon}/2$ with $\kappa=0,1,2,\ldots$ The force $F^*_{ab}(r)$ is obtained by extrapolating the data at the contact value $s=\sigma_{1b}$.

\subsubsection{Multicomponent solvent}
In a multicomponent solvent, we  have
$\rho^{(r)}(\qq)=\sum_\mu\rho^{(r)}_\mu(\qq)$ with $\rho^{(r)}_\mu(\qq)=\langle\sum_i\delta_{\mu_i,\mu}\delta(\qq-\rr_i)\rangle$, where the Greek index stands for the species, the Roman index stands for the particle label, and $\mu_i$ denotes the species of particle $i$. The depletion
force is now given by
\beq
\beta F_{ab}(r)=-\sum_\mu
\sigma_{\mu b}^2\int d\Omega_s\,\cos\theta_s\rho^{(r)}_\mu({\rr_b}+\sigma_{\mu b}\widehat{\mathbf{s}}).
\eeq

The output from the MC simulations are the functions
\beq \label{Imu}
I_\mu^{(r)}(s)=\left\langle\sum_i \delta_{\mu,\mu_i}
\frac{{\Pi_{s-\frac{\epsilon}{2},s+\frac{\epsilon}{2}}(s_i)}\cos\theta_{s_i}}{{\epsilon}}
\right\rangle,
\eeq
calculated at $s=s_\kappa=\sigma_{\mu b}+(2\kappa+1){\epsilon}/2$ with
$\kappa=0,1,2,\ldots$, so that we  now have
\beq \label{dforce}
F^*_{ab}(r)=\sigma_1\beta F_{ab}(r)=
-\sigma_1\sum_\mu I_\mu^{(r)}(\sigma_{\mu b}).
\eeq

\subsection{Rational-function approximation}

Within the RFA\cite{YHS96,YSH08,FS11,FS13} one  explicitly obtains
the Laplace transform $G_{ab}(s)$ of $rg_{ab}(r)$ in the solute
infinite-dilution limit ($x_a\to 0$ and $x_b\to 0$) of a quaternary
mixture where the solvent is made of  species $1$ and $2$ and the
solute is made of species $a= 3$ and $b= 4$. Then, from
Eq.\ (\ref{rfa-force}) we  have
\bal \label{rfa-force-1}
\beta F_{ab}(r)=&\frac{[rg_{ab}(r)]^\prime}{rg_{ab}(r)}-\frac{1}{r}\non
=&
\frac{\CL^{-1}[sG_{ab}(s)-e^{-\sigma_{ab}s}\sigma_{ab}g_{ab}(\sigma_{ab}^+)]}
{\CL^{-1}[G_{ab}(s)]}-\frac{1}{r},
\eal
where $\CL^{-1}$ stands for an inverse Laplace transform. In this
equation it is understood that $r>\sigma_{ab}$ since the force is of
course singular in the region $0\le r\le\sigma_{ab}$. Thus, {given that $\CL^{-1}[e^{-\sigma_{ab}s}]=\delta(r-\sigma_{ab})$}, we may
rewrite
\beq
\beta F_{ab}(r)=\frac{\CL^{-1}[sG_{ab}(s)]}{\CL^{-1}[G_{ab}(s)]}
-\frac{1}{r},\quad r>\sigma_{ab}.
\eeq

As discussed in Ref.\ \onlinecite{FS11}, the {RFA} inverse Laplace
transforms {for NAHS mixtures} could in principle present a spurious behavior in the shell
$\min(\sigma_{ab},\tau_{ab})\le r\le\max(\sigma_{ab},\tau_{ab})$,
where $\tau_{ab}$ is the minimum of the list of values
$\sigma_{bk}-(\sigma_k-\sigma_a)/2$ ($k=1$--$4$) that are different
from $\sigma_{ab}$. In our case, however, since the solute-solvent interaction
is additive, we have $\sigma_{bk}-(\sigma_k-\sigma_a)/2=\sigma_{ab}$
for all $k$, so that $\tau_{ab}=\sigma_{ab}$ and the spurious behavior
vanishes.

In the limit $\sigma_b/\sigma_a\to\infty$ the solute sphere $b$ is felt as a
planar hard wall by both a solvent particle and by the solute particle
$a$. Before taking the limit we introduce the shifted radial
distribution function $\gamma_{ab}(D)=g_{ab}(D+\sigma_{ab})$ for a
surface-to-surface distance $D\ge 0$. In Laplace space,
\beq
G_{ab}(s)=e^{-\sigma_{ab}s}[\sigma_{ab}\Gamma_{ab}(s)-\Gamma^\prime_{ab}(s)],
\label{GGamma}
\eeq
where $\Gamma_{ab}(s)$ is the Laplace transform of $\gamma_{ab}(D)$
and $\Gamma_{ab}^\prime(s)=d\Gamma_{ab}(s)/ds$. In the wall limit, Eq.\ \eqref{GGamma} yields
\beq
\Gamma_{aw}(s)=\lim_{\sigma_b/\sigma_a\to\infty}\frac{2}{\sigma_b}e^{\sigma_{ab}s}G_{ab}(s).
\eeq
The corresponding expression for the depletion force is
\bal
\beta F_{aw}(D)=&\frac{\gamma^\prime_{aw}(D)}{\gamma_{aw}(D)}=
\frac{\CL^{-1}[s\Gamma_{aw}(s){-\gamma_{aw}(0)}]}{\CL^{-1}[\Gamma_{aw}(s)]}\non
=&{\frac{\CL^{-1}[s\Gamma_{aw}(s)]}{\CL^{-1}[\Gamma_{aw}(s)]}},
\eal
{where in the last step we have taken into account that $D>0$ and thus the term coming from $\CL^{-1}[1]=\delta(D)$  can be ignored.}

Appendix \ref{app:idl} gives some  details on how to carry
out the solute infinite-dilution limit analytically, while Appendix
\ref{app:wl} shows how to subsequently carry out the wall limit.
Once $G_{ab}(s)$ and $\Gamma_{aw}(s)$ are known, the inverse Laplace transforms may be carried out numerically following the recipe of Ref.\ \onlinecite{AW92}.
When the solvent nonadditivity is switched off ($\Delta=0$) our RFA approach
reduces to the usual PY approximation.\cite{FS11,YSH08}

{The RFA for NAHS systems  inherits from the PY approximation for AHS fluids the possibility of yielding nonphysical results near contact for the big-big correlation function in the case of strongly asymmetric mixtures.\cite{HL86,JBH91,J96} As proposed by Henderson,\cite{H88} a simple and convenient way of circumventing this difficulty consists in the replacement $g\to\exp(g-1)$. Thus, in order to  correct the breakdown of the theory  near
solute contact, we have also considered an ``exponential'' RFA (exp-RFA)
approximation where\cite{LTJ09}}
\beq \label{exp-RFA}
g_{ab}^\text{exp-RFA}(r)=\exp\left[g_{ab}^\text{RFA}(r)-1\right].
\eeq

\section{Simulation details}
\label{sec:mc}

\begin{figure}[tbp]
\includegraphics[width=8cm]{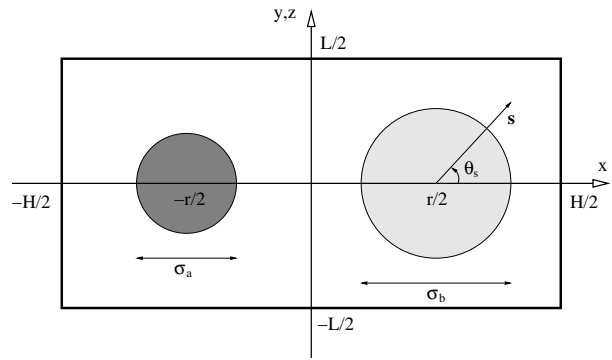}
\caption{Schematic simulation arrangement. The simulation box is the
  parallelepiped $-H/2<x<H/2$, $-L/2<y<L/2$, $-L/2<z<L/2$ with periodic
  boundary conditions. $H$ and $L$ are chosen large enough so as to have a solvent
  density exhibiting a bulk-like plateau away from the two solute  spheres.}
\label{fig:box}
\end{figure}

We performed  canonical MC simulations in a parallelepipedal box
($-H/2<x<H/2$, $-L/2<y<L/2$, $-L/2<z<L/2$) with periodic boundary
conditions. The two solute spheres $a$ and $b$ are fixed in space, centered at $(-r/2,0,0)$ and $(r/2,0,0)$, respectively, as
shown in Fig.\ \ref{fig:box}. The solvent is in  general a binary NAHS mixture, but we will always assume additivity between the solute and the solvent. According to the Metropolis algorithm,\cite{KW86} a solvent particle move is rejected
whenever it overlaps with another solvent particle or with any of the two
solute spheres. The maximum random particle displacement was chosen so
as to have acceptance ratios close to $50\%$. During the run we
measured the shell integrals $I_\mu^{(r)}(s)$ of Eq.\ (\ref{Imu}) and the
local solvent density.  We chose $H$ and $L$ large enough so that
away from the two solute spheres the local solvent density shows a
bulk-like plateau and thus {the  solvent density in a cubic cell of side $\ell$ centered at $(x,y,z)=(-H/2,L/2,L/2)$ can be accepted as a good estimate of the bulk density $\rho$.}

\begin{figure}[tbp]
\includegraphics[width=8cm]{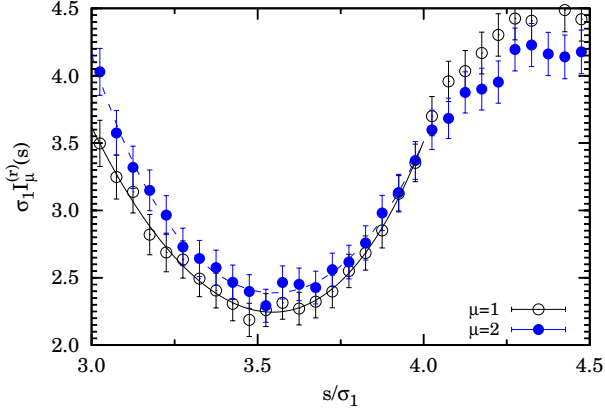}
\caption{ {Shell integrals
  $I_\mu^{(r)}(s)$  at $r/\sigma_1=5$ for the case $x_1=\frac{1}{2}$, $\sigma_2/\sigma_1=1$, $\Delta=0$,
  $\sigma_b/\sigma_a=1$, $\sigma_a/\sigma_1=5$. Here, $H/\sigma_1=18$, $L/\sigma_1=12$, $N=1134$. The bulk packing fraction  is  $\eta\approx 0.239(5)$ {and the simulation time was $\tau=4\times 10^5 N$
  single particle moves.} The lines are least-square quartic fits on the interval $3\leq s/\sigma_1\leq 4$ used to extrapolate $I_\mu^{(r)}(s)$ at contact ($s/\sigma_1=\sigma_{1b}/\sigma_1=3$). The estimated force is
  then found to be $F_{ab}^*(r)=-\sigma_1 \left[I_1^{(r)}(\sigma_{1b})+I_2^{(r)}(\sigma_{2b})\right]\approx -7.78(8)$.} This case is close to the one in Fig.\ 6b of
  Ref.\ \onlinecite{DAS97}.  }
\label{fig:fit}
\end{figure}

 A typical output for the shell integrals from
a single simulation is shown in Fig.\ \ref{fig:fit}. The uncertainty on
each measured value at a given $s$ is determined as
$\sqrt{\sigma_v^2 K/\tau}$ where $\tau$ is the number of single particle
moves, $\sigma_v^2$ is the variance of the measures during the run, and
$K$ is an estimate of the correlation time of the sequence of
measurements assumed as independent from $s$. In order to determine
the depletion force
according to Eq.\ (\ref{dforce}) we need to find the contact values
$I_\mu^{(r)}(\sigma_{\mu b})$. We do this with a  least-square quartic fit
of the shell integrals near contact, as shown in Fig.\ \ref{fig:fit}.
{Since the solvent binary mixture for the choice of the model parameters in Fig.\ \ref{fig:fit} reduces to a one-component  system,  no partial demixing is possible, so that the  $1\leftrightarrow 2$ symmetry implies the consequent equality of the two  shell integrals. This is reasonably well satisfied within
  the error estimates. The slight asymmetry observed in Fig.\ \ref{fig:fit}
favors one species or the other, in different runs, with equal probabilities.}

In the study of the wall limit $\sigma_b/\sigma_a\to\infty$, we removed the
periodic boundary conditions along the $x$ direction and placed a hard wall
at {$x= -H/2$} and another one at {$x= H/2$}, rejecting solvent-particle
moves  producing an overlap with the walls. The
solute sphere $a$ was placed on the $x$ axis at $x=-H/2+D+\sigma_a/2$ and
 the depletion force felt by the solute impurity
$\widehat{\xx}F_{wa}(D)$ was calculated as a function of $D>0$. {The solvent bulk density  was evaluated  in a cubic cell of side $\ell$ centered at $(x,y,z)=(D/2+\sigma_a/2,L/2,L/2)$.}

{One can take into account the volume excluded to the solvent particles by the solutes to define a (nominal) average packing fraction $\overline{\eta}=\overline{\eta}_1+\overline{\eta}_2$, where}
\beq
{\overline{\eta}_\mu=\frac{\frac{\pi}{6} N x_\mu \sigma_\mu^3}{HL^2-\frac{\pi}{6}\left(\sigma_{\mu a}^3+\sigma_{\mu b}^3\right)}}
\label{etamu}
\eeq
{if $\sigma_b/\sigma_a=\text{finite}$ and}
\beq
{\overline{\eta}_\mu=\frac{\frac{\pi}{6} N x_\mu \sigma_\mu^3}{(H-\sigma_\mu)L^2-\frac{\pi}{6}\sigma_{\mu a}^3}}
\label{etamuW}
\eeq
{if $\sigma_b/\sigma_a=\infty$.}

In all the cases presented in Sec.\ \ref{sec:results} we took $N=500$ solvent particles, box sides $H/\sigma_1=18$, $L/\sigma_1=12$, a number $\tau=1.4\times 10^6 N$ of single particle moves, and a discretization step ${\epsilon}/\sigma_1=0.05$. {The side of the cell employed to evaluate the bulk density was  $\ell=\sigma_1$.}

\section{Results}
\label{sec:results}

\begin{figure}[t]
\hspace{-3cm}
\includegraphics[width=4.5cm]{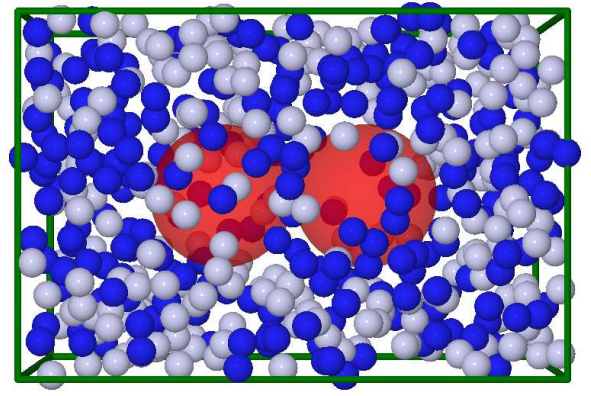}
\caption{ {Snapshot of an equilibrated MC configuration of system \sz}.  The solutes are the two big red spheres while the
  solvent binary mixture is made of small light and dark blue spheres.}
\label{fig:snap_S}
\end{figure}

\begin{table}
   \caption{{Values of the parameters defining the 12  systems considered in this work.}}\label{tab0}
\begin{ruledtabular}
\begin{tabular}{lccccrc}
Label&$\sigma_b/\sigma_a$& $\sigma_a/\sigma_1$&$x_1$&$\sigma_2/\sigma_1$ &$\Delta$ &$\overline{\eta}$ \\  \hline
\sz&$1$&$5$&$\frac{1}{2}$&$1$&$0$&$0.1021$\\
\spl&&&&&$\frac{1}{4}$&\\
\sm&&&&&$-\frac{1}{4}$&\\
\az&$1$&$5$&$\frac{193}{250}$&$\frac{3}{2}$&$0$&$0.1576$\\
\aplus&&&&&$\frac{1}{5}$&\\
\am&&&&&$-\frac{1}{5}$&\\
\wsz&$\infty$&$5$&$\frac{1}{2}$&$1$&$0$&$0.1076$\\
\wspl&&&&&$\frac{1}{4}$&\\
\wsm&&&&&$-\frac{1}{4}$&\\
\waz&$\infty$&$5$&$\frac{193}{250}$&$\frac{3}{2}$&$0$&$0.1685$\\
\wapl&&&&&$\frac{1}{5}$&\\
\wam&&&&&$-\frac{1}{5}$&\\
     \end{tabular}
 \end{ruledtabular}
 \end{table}

{In this section, we present our results for  four representative classes of systems: two
symmetric solute impurities in a symmetric (class S) or asymmetric (class A) solvent, and
a planar wall and a solute impurity in a symmetric (class wS) or asymmetric (class wA)
solvent. For each class, we have considered three solvent nonadditivities: zero, positive, and negative. This will allow us to assess the effect of solvent nonadditivity  on the depletion
force between the impurity particles or between the impurity and the wall. The RFA predictions will be compared with our MC
simulations.
The parameters characterizing the 12 different systems are given in Table \ref{tab0}. The last column gives the {average} packing fraction  $\overline{\eta}=\overline{\eta}_1+\overline{\eta}_2$ defined by Eqs.\ \eqref{etamu} (solute-solute systems) and \eqref{etamuW}  (wall-solute systems). In the asymmetric-solvent cases ($\sigma_2/\sigma_1=\frac{3}{2}$) the value of the mole fraction ($x_1=\frac{193}{250}$) has been chosen such that both species occupy practically equal volumes ($x_1\sigma_1^3/x_2\sigma_2^3=1.003$).}

{As an illustration, Fig.\ \ref{fig:snap_S}
shows a snapshot of an equilibrated MC configuration of system {\sz} with the two identical solute particles at contact.}

\subsection{Symmetric solvent and symmetric solute impurities}
\label{sub:s&s}

We first consider  a symmetric $1\leftrightarrow 2$ solvent (systems \sz, \spl, \sm). In general, for
positive nonadditivity ($\Delta>0$) and sufficiently high densities, the solvent may undergo demixing,\cite{SHY10} so that in the simulation we would get $I_1^{(r)}(s)\neq
I_2^{(r)}(s)$ by spontaneous symmetry breaking. On the other hand, if, at a given density,   the positive nonadditivity is not too large, the solvent will be in a mixed state and
 the equality of the two shell integrals is expected. However, we found
that, even in  states with a mixed solvent in the bulk,
the solvent may be partially demixed  in the  region between the two solute particles because of  density
compression effects.\cite{A89} This may  be responsible for an
asymmetry in the two shell integrals, which is expected to be maximal
near a surface-to-surface distance of the two solute impurities equal
to one solvent diameter. In order to avoid this effect, we chose a
sufficiently small value for the nonadditivity (system \spl).

\begin{table*}[tbp]
\caption{{MC results for the symmetric cases \sz, \spl,  \sm, and the asymmetric cases \az, \aplus,  {\am} (see Table \protect\ref{tab0}). $D$ is the surface-to-surface separation between the two solutes and
  $\eta$ is the bulk packing fraction of the solvent.}}
\label{tab:S}
\begin{ruledtabular}
\begin{tabular}{ccccccccccccc}
&\multicolumn{2}{c}{\sz}&\multicolumn{2}{c}{\spl}&\multicolumn{2}{c}{\sm}&\multicolumn{2}{c}{\az}&\multicolumn{2}{c}{\aplus}&\multicolumn{2}{c}{\am} \\
\cline{2-3} \cline{4-5} \cline{6-7} \cline{8-9} \cline{10-11} \cline{12-13}
$D/\sigma_1$ & $F^*_{ab}$ & $\eta$ & $F^*_{ab}$ & $\eta$ & $F^*_{ab}$ & $\eta$ & $F^*_{ab}$ & $\eta$ & $F^*_{ab}$ & $\eta$ & $F^*_{ab}$ & $\eta$\\
\hline
0.00&$-$2.35(3)&0.109(1)&$-$2.59(2)&0.108(1)&$-$2.22(2)&0.110(1)&$-$3.09(3)&0.167(1)&$-$3.39(3)&0.169(1)&$-$2.86(2)&0.168(1)\\
0.25&$-$1.71(2)&0.109(1)&$-$1.73(3)&0.109(1)&$-$1.59(2)&0.110(1)&$-$2.26(2)&0.166(1)&$-$2.43(4)&0.169(1)&$-$2.23(3)&0.171(1)\\
0.50&$-$1.03(2)&0.109(1)&$-$0.93(3)&0.109(1)&$-$1.01(3)&0.110(1)&$-$1.40(3)&0.168(1)&$-$1.33(3)&0.170(1)&$-$1.41(2)&0.169(1)\\
0.75&$-$0.30(3)&0.109(1)&$-$0.00(2)&0.109(1)&$-$0.40(2)&0.110(1)&$-$0.56(3)&0.169(1)&$-$0.24(3)&0.169(1)&$-$0.68(3)&0.170(1)\\
0.84&$-$0.03(2)&0.109(1)&~~0.25(3)&0.108(1)&$-$0.12(3)&0.110(1)&$-$0.22(3)&0.168(1)&~~0.13(3)&0.168(1)&$-$0.35(3)&0.169(1)\\
0.92&~~0.26(2)&0.109(1)&~~0.49(3)&0.108(1)&~~0.06(3)&0.110(1)&~~0.10(2)&0.168(1)&~~0.55(3)&0.168(1)&$-$0.10(3)&0.170(1)\\
1.00&~~0.36(3)&0.109(1)&~~0.66(3)&0.109(1)&~~0.21(3)&0.109(1)&~~0.45(2)&0.171(1)&~~0.95(4)&0.168(1)&~~0.17(2)&0.169(1)\\
1.08&~~0.32(4)&0.110(1)&~~0.63(4)&0.109(1)&~~0.27(3)&0.110(1)&~~0.50(4)&0.169(1)&~~0.76(6)&0.171(1)&~~0.15(4)&0.170(1)\\
1.16&~~0.17(4)&0.109(1)&~~0.28(3)&0.108(1)&~~0.08(3)&0.110(1)&~~0.21(4)&0.169(1)&~~0.36(5)&0.165(1)&~~0.03(3)&0.170(1)\\
1.25&$-$0.02(4)&0.109(1)&~~0.01(4)&0.108(1)&$-$0.02(3)&0.110(1)&~~0.07(5)&0.168(1)&~~0.23(4)&0.170(1)&$-$0.07(3)&0.170(1)\\
1.50&$-$0.01(4)&0.110(1)&~~0.02(4)&0.109(1)&$-$0.03(3)&0.110(1)&~~0.22(4)&0.169(1)&~~0.48(5)&0.166(1)&~~0.14(4)&0.170(1)\\
1.75&~~0.01(2)&0.109(1)&$-$0.13(4)&0.109(1)&$-$0.05(3)&0.109(1)&$-$0.07(4)&0.169(1)&$-$0.20(5)&0.168(1)&~~0.01(4)&0.170(1)\\
2.00&~~0.01(3)&0.109(1)&$-$0.11(3)&0.109(1)&$-$0.00(2)&0.110(1)&$-$0.10(3)&0.169(1)&$-$0.04(3)&0.168(1)&$-$0.02(3)&0.170(1)\\
2.25&$-$0.03(3)&0.109(1)&$-$0.06(3)&0.108(1)&$-$0.02(3)&0.110(1)&$-$0.04(3)&0.170(1)&$-$0.12(4)&0.171(1)&$-$0.06(3)&0.171(1)\\
2.50&~~0.02(2)&0.109(1)&~~0.04(2)&0.109(1)&$-$0.00(2)&0.109(1)&$-$0.00(3)&0.169(1)&~~0.09(3)&0.167(1)&$-$0.02(2)&0.169(1)\\
\end{tabular}
\end{ruledtabular}
\end{table*}

%
\begin{figure}[tbp]
\includegraphics[width=8cm]{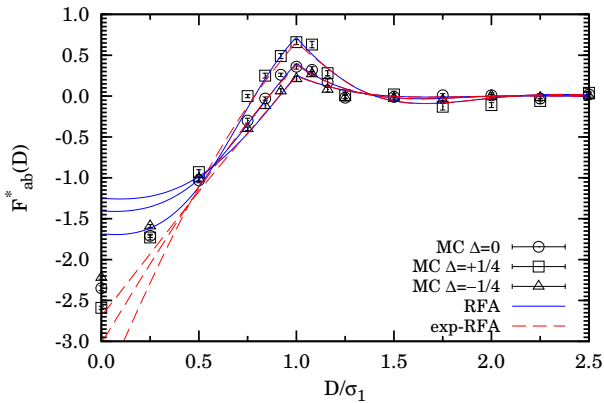}
\caption{ {Depletion force between two identical big
  hard spheres immersed in a solvent binary mixture of small
  hard spheres,
  as a function of their surface-to-surface separation, for systems \sz, \spl, and {\sm} (see Table \protect\ref{tab0}). The bulk packing fraction used to obtain the (exp-)RFA results  was taken
  as $\eta=0.109$ in all cases. The MC results are the ones of Table
  \ref{tab:S}.}}
\label{fig:S}
\end{figure}

The first columns of Table \ref{tab:S} present the simulation results for the depletion force and for the bulk packing fraction of systems \sz, \spl, and {\sm} as functions of the surface-to surface distance $D=r-\sigma_{ab}$.
{We observe that the bulk packing fraction is weakly dependent on $D$ and on $\Delta$, being slightly larger than the average value $\overline{\eta}$.}

The MC results for the depletion force are compared with the semi-analytical RFA predictions in Fig.\ \ref{fig:S}. We recall that the RFA theory reduces to the PY theory in the additive case ($\Delta=0$), so  the middle solid and dashed lines  in Fig.\ \ref{fig:S} actually represent the PY and exp-PY predictions, respectively.  As we can see, those curves for the additive system {\sz}  agree quite well with the simulation data at and beyond a surface-to-surface
separation between the two solute impurities equal to half the solvent diameter, $D\gtrsim \sigma_1/2$. In that region, our RFA theory successfully accounts for the influence of the solvent nonadditivity on the depletion force. A specially  good agreement is observed at $D=\sigma_1$, where the theory predicts a kink
in the force stemming from the first spatial derivative of the
solute-solute radial distribution function. On the other hand, a less
satisfactory result is observed near contact  of the impurities ($D<\sigma_1/2$), where both the PY (system \sz) and the RFA (systems {\spl} and \sm) theories exhibit an artificial upward bending of the curves (instead of the correct quasilinear behavior), implying a force less attractive than it should be. This is, at least qualitatively, corrected by the exp-PY and exp-RFA versions of the theories. Another possible correction could be to develop the
second-order RFA,\cite{HYS08} which is known to work well in
the additive solvent case.\cite{YSH08}

The positive nonadditivity enhances the depletion force and
the negative nonadditivity inhibits it.
These trends for the effect of the solvent nonadditivity on the
depletion force could be expected from the following simple argument. To first order in
density, the bulk compressibility factor of the solvent
is $1+B_2\rho$,
with $B_2=(2\pi/3)\sum_{i,j}x_ix_j\sigma_{ij}^3$ being the second virial
coefficient. Therefore, in the low-density regime, one would expect the NAHS solvent with a packing fraction $\eta$ to behave
similarly to an effective AHS solvent with an effective packing fraction
\beq
\eta_{\text{eff}}=\eta\frac{\sum_{i,j}x_ix_j\sigma_{ij}^3}
{\sum_{i,j}x_ix_j\left[
({\sigma_i+\sigma_j})/{2}\right]^3}.
\label{etaeff}
\eeq
Thus, introducing a positive nonadditivity in the solvent is qualitatively analogous to increasing its density, which in turn produces an enhancement of the solute-solute depletion force. Of course, a negative nonadditivity produces the opposite effect.

{\subsection{Asymmetric solvent and symmetric solute impurities}}
\label{sub:a&s}
Next, we consider the asymmetric-solvent systems \az, \aplus, and \am. In
those cases  the two shell integrals are obviously different, i.e., $I_1^{(r)}(s)\neq I_2^{(r)}(s)$. As before, we want to measure the effect on the depletion force of adding a certain
nonadditivity to the solvent.

\begin{figure}[tbp]
\includegraphics[width=8cm]{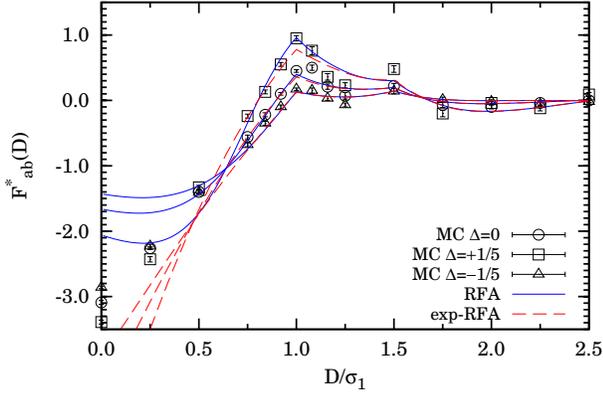}
\caption{
{Depletion force between two identical big
  hard spheres immersed in a solvent binary mixture of small
  hard spheres,
  as a function of their surface-to-surface separation, for systems \az, \aplus, and {\am} (see Table \protect\ref{tab0}). The bulk packing fraction used to obtain the (exp-)RFA results  was taken
  as $\eta=0.170$ in all cases. The MC results are the ones of Table
  \ref{tab:S}.}}
\label{fig:A}
\end{figure}

{The MC values for the depletion force and the bulk packing fraction are given in Table \ref{tab:S}. As in the symmetric-solvent cases, the bulk packing fractions are slightly larger than the nominal average values, but now the influence of the solute-solute separation on the bulk values is more pronounced.}

{Figure \ref{fig:A} compares the MC and RFA results for systems \az, \aplus, and \am.  As in the symmetric case, RFA=PY for the AHS solvent ($\Delta=0$).} Now, in addition to a kink in the depletion force at $D=\sigma_1$, the RFA predicts a second kink at
$D=\sigma_2$, with smooth oscillations around zero beyond that point.
{Again, the variation of the depletion force with distance increases (decreases) if a positive (negative) nonadditivity is included, as expected from the argument behind Eq.\ \eqref{etaeff}.} Analogous to Fig.\  \ref{fig:S}, a reasonable agreement between our theoretical
approximation and the simulation results is observed for $D\gtrsim\sigma_1/2$, but the agreement breaks down  when the solutes are near  contact. On the other hand,
the exp-RFA approximation has the correct linear behavior near contact,
even if it underestimates the contact values. {Note also that, while RFA and exp-RFA are practically indistinguishable for $D\gtrsim \sigma_1/2$ in Fig.\  \ref{fig:S}, both approximations are slightly different in the region near the kink at $D=\sigma_1$ in the case \aplus, RFA being more accurate than exp-RFA.}

\subsection{A wall and one solute impurity in a symmetric solvent}
\label{sub:s&w}

\begin{table*}[tbp]
\caption{{MC results for the symmetric cases \wsz, \wspl,  \wsm, and the asymmetric cases \waz, \wapl,  {\wam} (see Table \protect\ref{tab0}). $D$ is the surface-to-surface separation between the wall and the  solute sphere and
  $\eta$ is the bulk packing fraction of the solvent.}}
\label{tab:WS}
\begin{ruledtabular}
\begin{tabular}{ccccccccccccc}
&\multicolumn{2}{c}{\wsz}&\multicolumn{2}{c}{\wspl}&\multicolumn{2}{c}{\wsm}&\multicolumn{2}{c}{\waz}&\multicolumn{2}{c}{\wapl}&\multicolumn{2}{c}{\wam} \\
\cline{2-3} \cline{4-5} \cline{6-7} \cline{8-9} \cline{10-11} \cline{12-13}
$D/\sigma_1$ & $F^*_{aw}$ & $\eta$ & $F^*_{aw}$ & $\eta$ & $F^*_{aw}$ & $\eta$ & $F^*_{aw}$ & $\eta$ & $F^*_{aw}$ & $\eta$ & $F^*_{aw}$ & $\eta$\\
\hline
0.00&$-$4.44(3)&0.110(1)&$-$4.68(3)&0.108(1)&$-$4.10(2)&0.111(1)&$-$5.73(2)&0.172(1)&$-$6.20(2)&0.169(1)&$-$5.20(2)&0.170(1)\\
0.25&$-$3.29(2)&0.109(1)&$-$3.30(2)&0.108(1)&$-$3.11(2)&0.110(1)&$-$4.34(3)&0.168(1)&$-$4.49(3)&0.168(1)&$-$4.02(1)&0.172(1)\\
0.50&$-$1.99(2)&0.109(1)&$-$1.82(3)&0.108(1)&$-$2.01(2)&0.111(1)&$-$2.70(2)&0.171(1)&$-$2.55(3)&0.165(1)&$-$2.69(2)&0.171(1)\\
0.75&$-$0.66(2)&0.109(1)&$-$0.11(3)&0.108(1)&$-$0.79(3)&0.110(1)&$-$1.03(3)&0.170(1)&$-$0.32(4)&0.168(1)&$-$1.26(3)&0.172(1)\\
0.84&$-$0.06(2)&0.110(1)&~~0.60(3)&0.108(1)&$-$0.38(3)&0.111(1)&$-$0.30(3)&0.170(1)&~~0.56(3)&0.170(1)&$-$0.72(3)&0.172(1)\\
0.92&~~0.42(3)&0.109(1)&~~1.06(3)&0.108(1)&~~0.05(2)&0.110(1)&~~0.28(3)&0.169(1)&~~1.24(3)&0.169(1)&$-$0.22(3)&0.172(1)\\
1.00&~~0.95(3)&0.110(1)&~~1.66(2)&0.108(1)&~~0.59(3)&0.110(1)&~~1.00(3)&0.168(1)&~~2.01(3)&0.167(1)&~~0.34(3)&0.171(1)\\
1.08&~~0.83(5)&0.109(1)&~~1.42(6)&0.108(1)&~~0.52(4)&0.110(1)&~~0.92(7)&0.168(1)&~~1.91(10)&0.165(1)&~~0.43(5)&0.172(1)\\
1.16&~~0.33(5)&0.109(1)&~~0.44(7)&0.108(1)&~~0.19(4)&0.110(1)&~~0.41(8)&0.168(1)&~~0.99(10)&0.168(1)&~~0.16(5)&0.172(1)\\
1.25&$-$0.06(5)&0.110(1)&~~0.04(7)&0.109(1)&$-$0.05(5)&0.110(1)&~~0.11(8)&0.169(1)&~~0.54(10)&0.167(1)&$-$0.04(7)&0.171(1)\\
1.50&~~0.08(6)&0.110(1)&$-$0.02(7)&0.109(1)&~~0.12(5)&0.110(1)&~~0.55(7)&0.171(1)&~~0.93(9)&0.165(1)&~~0.38(7)&0.172(1)\\
1.75&$-$0.06(5)&0.109(1)&$-$0.17(6)&0.109(1)&$-$0.07(4)&0.110(1)&$-$0.15(7)&0.170(1)&$-$0.56(8)&0.167(1)&$-$0.07(6)&0.171(1)\\
2.00&$-$0.00(2)&0.110(1)&$-$0.06(3)&0.108(1)&~~0.01(2)&0.110(1)&$-$0.04(5)&0.171(1)&$-$0.28(5)&0.167(1)&$-$0.02(3)&0.171(1)\\
2.25&$-$0.03(3)&0.109(1)&~~0.05(3)&0.108(1)&~~0.00(2)&0.110(1)&$-$0.05(4)&0.168(1)&$-$0.16(4)&0.167(1)&$-$0.00(3)&0.172(1)\\
2.50&~~0.05(2)&0.109(1)&~~0.03(3)&0.108(1)&$-$0.03(2)&0.111(1)&~~0.07(3)&0.170(1)&$-$0.01(3)&0.165(1)&~~0.01(2)&0.172(1)\\
\end{tabular}
\end{ruledtabular}
\end{table*}

We now explore the cases of extreme solute asymmetry in the limit
$\sigma_b/\sigma_a\to\infty$, where sphere $b$ becomes a planar hard wall.

We start  with the cases of a
symmetric solvent (systems \wsz, \wspl, and \wsm). {The MC data for the depletion force and the bulk packing fraction are listed in the first columns of Table \ref{tab:WS}.} Since the solvents in systems \wsz, \wspl, and {\wsm} are in the same bulk state (except for small changes of $\eta$) as in systems \sz, \spl, and {\sm}, respectively,  we can test the
Derjaguin approximation\cite{D34} $F_{aa}^*(D)\approx \frac{1}{2}{F_{aw}^*}(D)$. As can be seen from comparison of Tables \ref{tab:S} and \ref{tab:WS}, the Derjaguin approximation is rather
well satisfied in our simulations, even in the cases of  NAHS
solvents, {$\frac{1}{2}F_{aw}^*(D)$ being typically $1$--$10$\% smaller than $F_{aa}^*(D)$.}

\begin{figure}[tbp]
\includegraphics[width=8cm]{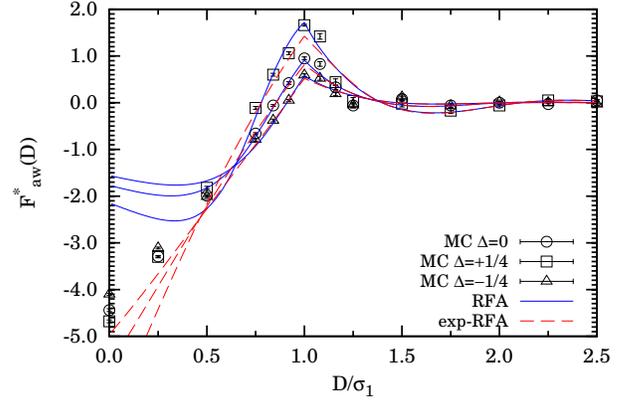}
\caption{
{Depletion force between a hard wall and a big
  hard sphere immersed in a solvent binary mixture of small
  hard spheres,
  as a function of their surface-to-surface separation, for systems \wsz, \wspl, and {\wsm} (see Table \protect\ref{tab0}). The bulk packing fraction used to obtain the (exp-)RFA results  was taken
  as $\eta=0.109$ in all cases. The MC results are the ones of Table
  \ref{tab:WS}.}}
\label{fig:WS}
\end{figure}

Theory and simulation are compared in Fig.\ \ref{fig:WS}. Not surprisingly, our RFA approximation (which is again equivalent to the PY approximation in the case $\Delta=0$)
performs quite well for $D\gtrsim \sigma_1/2$ but it {breaks} down near contact
between the wall and the solute spherical impurity, this effect being now more important than in the cases of two identical solutes  (Fig.\ \ref{fig:S}).
On the other hand, the exp-RFA approximation exhibits a better (quasilinear) behavior near contact, although it underestimates the contact values. {Also, analogous to what is observed in Fig.\ \ref{fig:A}, exp-RFA is less accurate than RFA near the kink at $D=\sigma_1$ when a positive nonadditivity is present.}

\subsection{A wall and one solute impurity in an asymmetric solvent}
\label{sub:a&w}
{To complete the picture, we finally consider the wall-solute force in a NAHS solvent (systems \waz, \wapl, and \wam). The corresponding MC data can be found in Table \ref{tab:WS}. The
Derjaguin approximation $F_{aa}^*(D)\approx \frac{1}{2}F_{aw}^*(D)$ is again
well satisfied, although the deviations are slightly larger than in the wS cases, $\frac{1}{2}F_{aw}^*(D)$ being about $4$--$10$\% smaller than $F_{aa}^*(D)$.}

\begin{figure}[tbp]
\includegraphics[width=8cm]{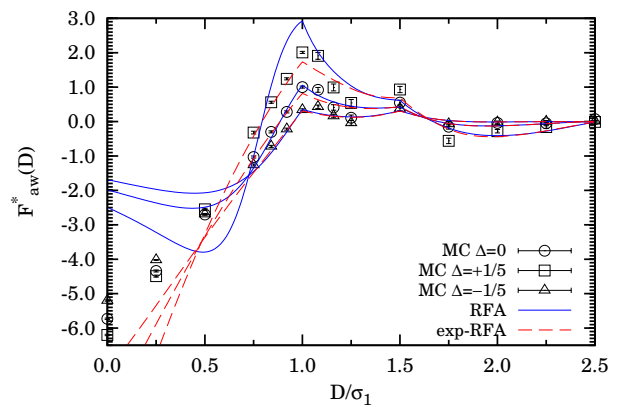}
\caption{
{Depletion force between a hard wall and a big
  hard sphere immersed in a solvent binary mixture of small
  hard spheres,
  as a function of their surface-to-surface separation, for systems \waz, \wapl, and {\wam} (see Table \protect\ref{tab0}). The bulk packing fraction used to obtain the (exp-)RFA results  was taken
  as $\eta=0.170$ in all cases. The MC results are the ones of Table
  \ref{tab:WS}.}
}
\label{fig:WA}
\end{figure}

As Fig.\ \ref{fig:WA} shows, in contrast to the cases \spl, \aplus, and \wspl plotted in Figs.\ \ref{fig:S}, \ref{fig:A}, and \ref{fig:WS}, respectively,  the RFA for a
positive nonadditivity (system \wapl) is not able to capture accurately the values of the depletion force in the region near the first kink
at $D=\sigma_1$, while the related approximation exp-RFA does. Moreover, the artificial upward bend of the PY curve ($\Delta=0$) and of the two  RFA curves ($\Delta=\pm \frac{1}{5}$) in the region $D\lesssim \sigma_1/2$ is much more dramatic than in Figs.\ \ref{fig:S}--\ref{fig:WS}. Again,  the exp-RFA lines tend to correct this behavior but they underestimate the contact values.

\section{Conclusions}
\label{sec:conclusions}

We have studied in this paper the mutual depletion force acting on two solute
hard spheres immersed in a solvent consisting in a binary NAHS mixture. We have employed two complementary tools: canonical MC simulations and the semi-analytical RFA (which is fully equivalent to the PY solution when the solvent nonadditivity is switched off). Four different settings have been considered:  two symmetric
solutes in
a symmetric and in an asymmetric solvent, and two extremely asymmetric
solutes (in the limit where one of the two spheres reduces to a planar
hard wall) again in a symmetric and in an asymmetric solvent. For each class of systems we have chosen three possibilities: zero nonadditivity, positive nonadditivity, and negative nonadditivity. In all the systems the solvent remained in a mixed state.

We have found that the RFA performs reasonably well in all cases for a
surface-to-surface distance $D$ greater than the radius of the
smallest solvent particles, except in the case {\wapl} of a wall with an
asymmetric solvent with positive nonadditivity, where the theory overestimates the height of the first kink. The approximation in
all cases breaks down at and near contact ($D=0$). To correct this, we
have also considered an exp-RFA, which shows the correct quasilinear behavior
near contact, even if it  is still not able to quantitatively capture the
contact values. The approximations correctly
predict kinks in the depletion force when $D$ equals any of the two solvent
diameters. Our results show how in all cases a positive solvent
nonadditivity enhances the depletion force whereas a negative one
inhibits it. Moreover, the Derjaguin approximation is well satisfied in our
simulations, even for the nonadditive solvent.

{As possible further developments of our study, we plan} to try to correct
the theoretical approximation near contact and to study the behavior of
the force as one approaches the demixing transition of the solvent on
the critical isochore.

\begin{acknowledgments}

R.F. acknowledges the hospitality of the University of
Extremadura in Badajoz, where the work was carried out, and the use of
the CINECA computational facilities under the ISCRA grant.
The research of A.S.  was supported by the Spanish Government through Grant No.\ FIS2010-16587 and  by the Junta de Extremadura (Spain) through Grant No.\ GR10158, both partially financed by FEDER funds.
\end{acknowledgments}

\appendix
\section{The solute infinite-dilution limit in the RFA}
\label{app:idl}

For convenience, we here use Roman indexes for the species instead of Greek
indexes as done in the main text.
In Ref.\ \onlinecite{FS11}, the following proposal for the
structural properties of an $n$-component NAHS fluid defined through the
Laplace transform $G_{ij}(s)$ of $rg_{ij}(r)$ was given:
\beq
G_{ij}(s)=s^{-2}\sum_{k=1}^n e^{-\sigma_{ik}s}L_{ik}(s)B_{kj}(s),
\label{Gij}
\eeq
with
\beq
\mathsf{B}^{-1}(s)= \mathsf{I}-\mathsf{A}(s),
\label{Bij}
\eeq
\beq
A_{ij}(s)= \frac{2\pi{{\rho}}
{{x}}_i}{s^3}\left[N_{ij}(s)e^{a_{ij}s}-L_{ij}(s)e^{-\sigma_{ij}s}\right],
\label{Qij}
\eeq
where $\mathsf{I}$ is the unit matrix,
\beq
{L}_{ij}(s)\equiv {L}_{ij}^\zero+{L}_{ij}^\one s,
\eeq
\beq
N_{ij}(s)\equiv L_{ij}^\zero\left(1-b_{ij} s+\frac{b_{ij}^2
  s^2}{2}\right)+L_{ij}^\one s\left(1-b_{ij} s\right),
\label{Nkj}
\eeq
\beq
b_{ij}\equiv
\sigma_{ij}+a_{ij}, \quad a_{ij}\equiv \frac{1}{2}(\sigma_i-\sigma_j).
\label{bij}
\eeq
Equations \eqref{Gij}--\eqref{Nkj} provide the explicit $s$-dependence
of the Laplace transform $G_{ij}(s)$, but it still remains to determine
the two sets of parameters $L_{ij}^\zero$ and $L_{ij}^\one$. This is
done by enforcing the physical requirements\cite{FS11}
$\lim_{s\to 0}s^2 G_{ij}(s)=1$ and $\lim_{s\to 0}s^{-1}\left[s^2
  G_{ij}(s)-1\right]=0$. The results are
\beq
L_{ij}^\zero=S_j,\quad
L_{ij}^\one=T_j+\sigma_{ij}S_j,
\label{Lij0}
\eeq
where
\beq
S_j\equiv\frac{1-\pi {{\rho}}\Psi_j}{\left(1-\pi
{{\rho}}\Lambda_j\right)\left(1-\pi
{{\rho}}\Psi_j\right)-\pi^2{{\rho}}^2\mu_{j|2,0}\Omega_j},
\label{28a}
\eeq
\beq
T_j\equiv\frac{\pi{{\rho}}\Omega_j}{\left(1-\pi {{\rho}}\Lambda_j\right)\left(1-\pi
{{\rho}}\Psi_j\right)-\pi^2{{\rho}}^2\mu_{j|2,0}\Omega_j},
\label{29a}
\eeq
\beq
\Lambda_j\equiv \mu_{j|2,1}-\frac{1}{3}\mu_{j|3,0},
\label{30a}
\eeq
\beq
\Psi_j\equiv \frac{2}{3}\mu_{j|3,0}-\mu_{j|2,1},
\label{31}
\eeq
\beq
\Omega_j\equiv \mu_{j|3,1}-\mu_{j|2,2}-\frac{1}{4}\mu_{j|4,0},
\label{32a}
\eeq
and we have called
\beq
\mu_{j|p,q}\equiv \sum_{k=1}^n {{x}}_k b_{kj}^p \sigma_{kj}^q.
\label{25}
\eeq

We now choose our  quaternary mixture ($n=4$) in such a way that the first
two species ($i=1$ and $i=2$) describe the solvent and the last two species
($i=3= a$ and $i= 4= b$) describe the solute. Then, in the infinite-dilution limit {$x_a\to 0$} and {$x_b\to 0$} we have that
\beq
\mathsf{B}^{-1}=\left(\begin{array}{cccc}
(\mathsf{B}^{-1})_{11} & (\mathsf{B}^{-1})_{12} & {-A_{1a}} & {-A_{1b}}\\
(\mathsf{B}^{-1})_{21} & (\mathsf{B}^{-1})_{22} & {-A_{2a}} & {-A_{2b}}\\
0                    & 0                    & 1      & 0\\
0                    & 0                    & 0      & 1\\
\end{array}\right),
\eeq
and thus
\beq
\mathsf{B}=\left(\begin{array}{cccc}
B_{11} & B_{12} & {C_{1a}}  & {C_{1b}}\\
B_{21} & B_{22} & {C_{2a}}  & {C_{2b}}\\
0     & 0     & 1      & 0\\
0     & 0     & 0      & 1\\
\end{array}\right),
\eeq
where
\beq
C_{ij}=\sum_{k=1}^2B_{ik}A_{kj},\quad i=1,2~~\mbox{and}~~j={a,b}.
\eeq
We have reduced the inversion of the original $4\times 4$ matrix
$\mathsf{B}^{-1}$ to the inversion of just the $2\times 2$
submatrix corresponding to   the solvent.

We then find
\beq \label{Gidl}
{s^2G_{ab}(s)=e^{-\sigma_{ab}s}L_{ab}(s)+
\sum_{k=1}^2e^{-\sigma_{ak}s}L_{ak}(s)C_{kb}(s)},
\eeq
where now $\mu_{j|p,q}= \sum_{k=1}^2 {{x}}_k b_{kj}^p \sigma_{kj}^q$.

\section{The wall limit in the RFA}
\label{app:wl}

Taking the limit  $\sigma_b\to\infty$, we find from Eq.\ \eqref{Gidl},
\bal
{\Gamma_{aw}(s)}=&{\lim_{\sigma_b\to\infty}\frac{2}{\sigma_b}e^{\sigma_{ab}s}G_{ab}(s)}\non
=&{\frac{2}{s^2}\left[\widetilde{L}_{aw}(s)+\sum_{k=1}^2L_{ak}(s)\widetilde{C}_{kw}(s)
\right]},
\eal
where
\beq
{\widetilde{L}_{aw}(s)\equiv\lim_{\sigma_b\to\infty}
\frac{L_{ab}(s)}{\sigma_b}},
\eeq
\beq
{\widetilde{C}_{kw}(s)\equiv\lim_{\sigma_b\to\infty}
\frac{e^{a_{kb}s}C_{kb}(s)}{\sigma_b}},\quad k=1,2.
\eeq

\bibliographystyle{apsrev}

\bibliography{D:/Dropbox/Public/bib_files/liquid}

\end{document}